# 14 New Light Curves and an Updated Ephemeris for the Hot Jupiter HAT-P-54 b

**Heather B. Hewitt**
*School of Earth and Space Exploration, Arizona State University, 781 E. Terrace Mall, Tempe, AZ 85287-6004; hbhewittt@asu.edu*

| | | |
|---|---|---|
| **Bradley Hutson** | **Atea Johnson** | **Elizabeth Quinlan** |
| **Michael Brockman** | **Chris Kight** | **Eva Randazzo** |
| **Elizabeth Catogni** | **Ryan A. Kilinski** | **Kellan Reagan** |
| **Rosemary Ferreira** | **Khatu Nguyen** | **Kinley Subers** |
| **Gary Fussell** | **Ty Perry** | **Federico R. Noguer** |

*School of Earth and Space Exploration, Arizona State University, 781 E. Terrace Mall, Tempe, AZ 85287-6004*

**Molly N. Simon**
*School of Earth and Space Exploration, Arizona State University, 781 E. Terrace Mall, Tempe, AZ 85287-6004*

**Robert T. Zellem**
*Jet Propulsion Laboratory, California Institute of Technology, 4800 Oak Grove Drive, Pasadena, CA 91109, and NASA Goddard Space Flight Center, Exoplanets and Stellar Astrophysics Laboratory (Code 667), Greenbelt, MD 20771*

**Abstract** Here we present an analysis of 14 transit light curves of the hot Jupiter HAT-P-54 b. Thirteen of our datasets were obtained with the 6-inch MicroObservatory telescope, Cecilia, and one was measured with the 61-inch Kuiper Telescope. We used the EXOplanet Transit Interpretation Code (EXOTIC) to reduce 49 datasets in order to update the planet's ephemeris to a mid-transit time of 2460216.95257 ± 0.00022 BJD_TBD and an updated orbital period of 3.79985363 ± 0.00000037 days. These results improve the mid-transit uncertainty by 70.27% from the most recent ephemeris update. The updated mid-transit time can help to ensure the efficient use of expensive, large ground- and space-based telescope missions in the future. This result demonstrates that amateur astronomers and citizen scientists can provide meaningful, cost-efficient, crowd-sourcing observations using ground-based telescopes to further refine current mid-transit times and orbital periods.

### 1. Introduction

To date, there have been over 5,000 confirmed exoplanets discovered, with over 7,000 candidates yet to be validated (NASA Exoplanet Archive 2023). While it is impractical to perform follow up using solely space- or large ground-based telescopes, which are generally expensive to operate and under strict observing schedules, large-scale citizen science projects, such as Exoplanet Watch[1] (Zellem *et al.* 2020), use crowd-sourced data to improve upon results from expensive space-based missions (e.g. Mizrachi *et al.* 2021; Hewitt *et al.* 2023a). Exoplanet Watch solicits the help of the public by utilizing small (oftentimes privately owned) telescopes to validate findings or improve the ephemerides of exoplanets for potential follow-up observation with large ground-based or space-based telescopes.

Over time, a planet's mid-transit time can become "stale" in which the uncertainty of orbital times grows, causing the uncertainty of future transit times to require additional observing time in order to capture the entire transit (Zellem *et al.* 2020). This can become problematic, as valuable time and money can be lost if a transit time is miscalculated and missed. Small instruments can be used to collect data and amateur astronomers can analyze these observations to produce light curves and update orbital parameters of the planet. Studies using large amounts of observations taken by small telescopes prove to be invaluable to further refine the published knowledge. Small robotic telescopes, used by groups like Exoplanet Watch, can measure the orbital parameters of 195 known exoplanets to 3σ, and the number of exoplanets that can be viewed by ground-based telescopes will expand as the number of discovered exoplanets from JWST, TESS, or ARIEL increases (Zellem *et al.* 2020). Undergraduate students in the Spring 2023 offering of an online research course (discussed below) partnered with Exoplanet Watch to update the mid-transit time and orbital period of HAT-P-54 b, a hot Jupiter exoplanet (a = 0.04 AU; P = 3.8 days; M = 0.760 ± 0.032 $M_{Jup}$; Bakos *et al.* 2015) orbiting a K-type star 442.93-ly from the Earth.

This study was accomplished at Arizona State University (ASU) in one of the first course-based undergraduate research experiences (CUREs) for online astronomy majors. This fifteen-week course, titled Exoplanet Research Experience, was developed to offer non-traditional, online learners the opportunity to participate in authentic research experiences. CUREs help to make research accessible to a more diverse learning population, including people with full-time jobs, parents, veterans, and persons with disabilities (Auchincloss *et al.* 2014; Hewitt *et al.* 2023b).

---
[1] https://exoplanets.nasa.gov/exoplanet-watch/about-exoplanet-watch/overview/

## 2. Observatory and observing conditions

Forty-eight observations were collected using the Cecilia telescope of MicroObservatory. Operated by the Harvard-Smithsonian Center for Astrophysics, MicroObservatory is a network of five automated 6-inch telescopes that provides an accessible avenue for students to make astronomy a more hands-on and interactive laboratory in an online environment. Most of our observations were taken with MicroObservatory's Cecilia telescope, which is located at the Fred Lawrence Whipple Observatory on Mt. Hopkins in Arizona.

Cecilia is a custom-built Maksutov-Newtonian with a 6-inch diameter mirror and a focal length of 560 mm. Its imaging sensor is a custom Kodak KAF1400 CCD with pixels 6.9 μm per side producing an image 0.94 × 0.72 degrees, using 2 × 2 pixel binning, giving an overall resolution of 5.0 arc-second/pixel (Sadler *et al.* 2001). All image series used from MicroObservatory had exposure times of 60 seconds using a clear filter.

The MicroObservatory telescope network's weather ratings rely on data provided by NOAA IR Satellite images (Observing with NASA (CFA Harvard 2023). MicroObservatory's weather rating ranges from 000 to 100, with 000 indicating a complete overcast, and 100 indicating a clear night sky. Weather ratings are determined through an automated process wherein software encircles the location of where the telescope is on the satellite image, then places the pixels within the circle on a scale from 000 to 100 (Sienkiewicz 2021). Although weather ratings provide a guide, the NOAA weather ratings are not always entirely accurate. Therefore, a more in-depth analysis of each night of data is required to determine the quality of the dataset. All nights of data used to determine the mid-transit timings had recorded weather scores greater than 90, with the exception of several dates in 2018 and 2019 when MicroObservatory weather rating parameters were unavailable.

An additional dataset was obtained using the Steward Observatory 61-inch Kuiper Telescope atop Mt. Bigelow, Arizona. The telescope uses the cryo-cooled Mont4K camera, which features 15 μm pixels per side, resulting in an image 580 arcsec x 580 arcsec (9.7 arcmin x 9.7 arcmin) at a resolution of 0.14 arcsec/pixel (Univ. Arizona 2023). The image series obtained from Kuiper were 10 seconds while using the clear filter and there was no evidence of imprecise tracking.

Weather data are not recorded in the same manner for the 61-inch Kuiper Telescope observation as they are for MicroObservatory, but it should be noted that on 2023-09-29 skies were clear, with a moon at 100% illumination and 94 degrees from target.

## 3. Data reduction

For the photometric evaluation of our data, we used Exoplanet Watch's software, the EXOPLANET TRANSIT INTERPRETATION CODE (EXOTIC; Zellem *et al.* 2020). EXOTIC is a PYTHON 3 pipeline designed for the analysis and interpretation of exoplanet transit observations. It processes data in the form of ".fits" image files to locate and track the host star over the course of a night in order to derive information such as the light curve and the planet's orbital parameters. EXOTIC can be run locally or on the Google Colaboratory (Colab). For our purposes, we chose to run EXOTIC on the Google Colab to preclude all of our CURE student members having to install PYTHON and its various libraries on their local machines.

After an initial run of the data, we identified a list of the strongest candidates for significant detections. To determine which light curves should be included in this study, we determined that a light curve with a detection significance equal to or greater than 3σ would be considered a significant detection.

We re-ran significant and borderline significant (>2.8σ;<3σ) detections through EXOTIC using a standardized method similar to that used in Hewitt *et al.* (2023a): we first identified and manually removed any "bad" images, which we defined as images with deficient telescope tracking or weather-related incidents (i.e., cloud cover) that significantly obstructed our view of the host or comparison star. Next, with no recommended comparison stars available from the American Association of Variable Star Observers (AAVSO) in the field of view, we chose comparison stars close to HAT-P-54 to ensure that they were affected similarly by systematic errors, such as air mass variations. To identify the set of the best comparison stars, we input multiple potential comparison stars and let EXOTIC select the one that best reduced the scatter of the residuals from each of the individual datasets. We then identified the top five best comparison stars from all of the analyses and re-analyzed each significant and borderline dataset with EXOTIC with these top five comparison stars (Figure 1).

With this standardized analysis method, we found that 14 of the original 48 MicroObservatory datasets and the single Kuiper telescope observation had statistically significant detections, which we present here.

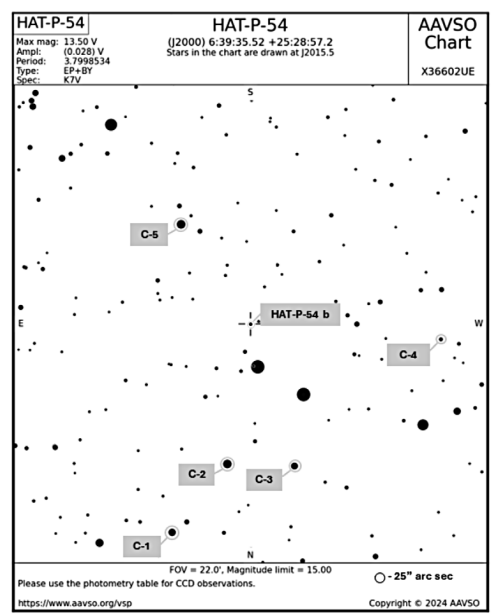

Figure 1. AAVSO VSP star chart for HAT-P-54. The field of view is 22' and the largest aperture used was 49.61". The comparison stars used in this study are labeled C-1 through C-5.

## 4. Data

We reduced 14 statistically significant light curves of HAT-P-54 b from our original 49 nights of data from MicroObservatory and the Kuiper telescope (Table 1, Figure 2). All 14 of the light curves used in this study are provided in Appendix A.

In order to calculate an updated mid-transit time and orbital period, we created an Observed–Calculated (O–C) plot, presented in Figure 3. The 14 mid-transit times analyzed in this study were included in the plot as well as the reported mid-transit time from the initial discovery paper (Bakos *et al.* 2015), four mid-transit times from Saha *et al.* (2021) that were fit by Ivshina and Winn (2022), and six additional mid-transit times from Ivshina and Winn (2022). Other studies in the NASA Exoplanet Archive (Bonomo *et al.* 2017; Kokori *et al.* 2022) were excluded from the creation of the O–C plot in an effort to keep the data consistent and only include those studies that derived the mid-transit time from a light curve. The values used from Bakos *et al.* (2015), Saha *et al.* (2021), and Ivshina and Winn (2022) are shown in Table 2 and the 14 values used from this study are shown in Table 1. For the creation of the O–C plot, we used an Exoplanet Watch notebook, "Exoplanet Ephemeris Fitting Tutorial," which allows for the generation of the O–C plot and posterior plot distribution and for the calculation of the mid-transit time and orbital period.[2] The mid-transit time from the most recent observation, that from the Kuiper telescope on 2023-09-29, and the most recently published period (3.7998529 ± 0.0000017 days; Ivshina and Winn 2022) were used as priors. We updated the mid-transit time to be 2460216.95257 ± 0.00022 BJD_TDB and the orbital period to be 3.79985363 ± 0.00000037 days. The O–C plot and the posterior plot distribution are presented in Figures 3 and 4, respectively.

Given that the point spread function (PSF) of our target star had an average full-width half-max of ~26", the light from nearby stars (located 24" and 32" away) could enter the aperture and dilute the transit signal (e.g., Crossfield *et al.* 2012; Bergfors *et al.* 2013; Stevenson *et al.* 2014; Ciardi *et al.* 2015; Collins *et al.* 2018; Zellem *et al.* 2020). However, this effect does not impact our measured mid-transit time and is therefore inconsequential for this work to update the ephemeris of HAT-P-54 b. Despite this potential dilution, we still see a 2-σ agreement between our derived transit depths (Table 1) and those previously reported in the literature (Bakos *et al.* 2015) and, as we discuss in detail in the Results section, we can improve upon the reported mid-transit ephemeris. Similarly, MicroObservatory's imprecise tracking during its 60-sec exposures (which manifests as oblate PSFs) can be tolerated, as the extraction and accurate measurement of the total number of photons measured during this time is the most important facet.

## 5. Results

To compare our updated mid-transit time uncertainty (listed above) to those found previously, we forward-propagated

Table 1. Mid-transit times and transit depths for the 14 significant detections calculated by EXOTIC.

| Date | Mid-Transit (BJD_TDB) (+2450000) | Mid-Transit Uncertainty (days) | Transit Depth $(R_p^2/R_s^2)$ (%) | Transit Depth Uncertainty |
|---|---|---|---|---|
| 2016-01-24 | 7412.6674 | 0.0032 | 0.0326 | 0.0059 |
| 2016-03-22 | 7469.6525 | 0.0029 | 0.0397 | 0.0034 |
| 2016-12-08 | 7731.8468 | 0.0014 | 0.0301 | 0.0039 |
| 2017-11-15 | 8073.8332 | 0.0013 | 0.0317 | 0.0060 |
| 2018-02-22 | 8172.6313 | 0.0016 | 0.0320 | 0.0044 |
| 2018-10-23 | 8415.8145 | 0.0030 | 0.0365 | 0.0099 |
| 2019-01-07 | 8491.8167 | 0.0032 | 0.0398 | 0.0066 |
| 2019-11-27 | 8814.8065 | 0.0034 | 0.0279 | 0.0049 |
| 2020-01-27 | 8875.6101 | 0.0036 | 0.0366 | 0.0053 |
| 2021-01-17 | 9232.7838 | 0.0062 | 0.0213 | 0.0060 |
| 2021-02-24 | 9270.8065 | 0.0058 | 0.0400 | 0.0100 |
| 2021-12-03 | 9551.9896 | 0.0030 | 0.0400 | 0.0037 |
| 2022-03-30 | 9669.7897 | 0.0061 | 0.0398 | 0.0088 |
| 2023-09-29* | 10216.9531 | 0.00046 | 0.02003 | 0.0005 |

*Denotes the 61" Kuiper observation.*

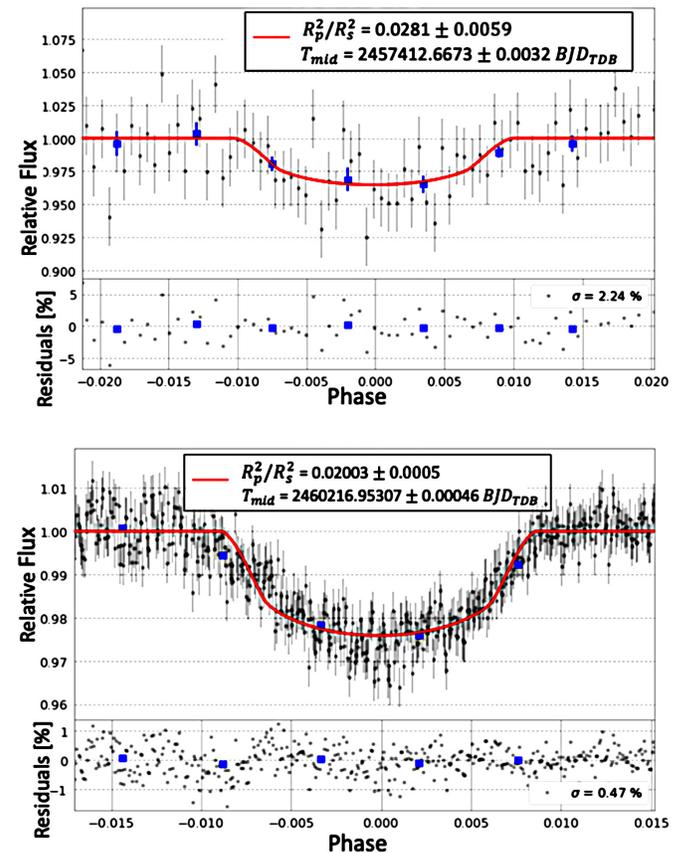

Figure 2. An example of a significant light curve from MicroObservatory on 2016-01-24 (top) and the light curve from the Kuiper telescope on 2023-09-29 (bottom). The gray data points represent data collected from each image of the data set. The blue data points represent an average from a set of binned data points and were used to fit the light curve.

---

[2] The ephemeris fitter can be found at https://colab.research.google.com/drive/1T5VT2gZ-ip6K6T9IXqMzQdSiEaf-UbJn?usp=sharing

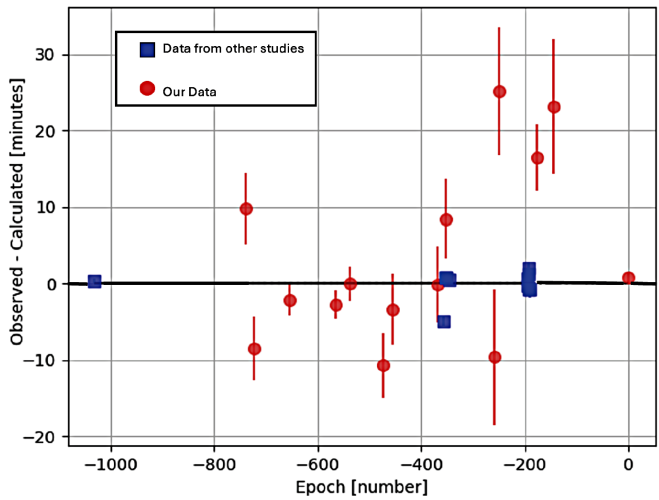

Figure 3. O–C plot for HAT-P-54 b using data from this study as well as previous observations. The priors used were $t_0$ = 2460216.95298 and p = 3.7998529.

Table 2. Prior solutions included in our combined O–C plot (Figure 4).

| Mid-Transit (BJD_TDB) (+2450000) | Mid-Transit Uncertainty (days) | Reference |
|---|---|---|
| 6299.30370 | 0.00024 | Bakos *et al.* (2015) |
| 8864.20126 | 0.00041 | Ivshina and Winn (2022) fit of Saha *et al.* (2021) |
| 8883.20451 | 0.000427 | Ivshina and Winn (2022) fit of Saha *et al.* (2021) |
| 8883.20456 | 0.000437 | Ivshina and Winn (2022) fit of Saha *et al.* (2021) |
| 8902.20360 | 0.000308 | Ivshina and Winn (2022) fit of Saha *et al.* (2021) |
| 9475.98097 | 0.0006831 | Ivshina and Winn (2022) |
| 9479.78137 | 0.0006766 | Ivshina and Winn (2022) |
| 9483.58175 | 0.0007356 | Ivshina and Winn (2022) |
| 9487.38019 | 0.0007031 | Ivshina and Winn (2022) |
| 9491.18192 | 0.0006023 | Ivshina and Winn (2022) |
| 9494.97987 | 0.0007512 | Ivshina and Winn (2022) |

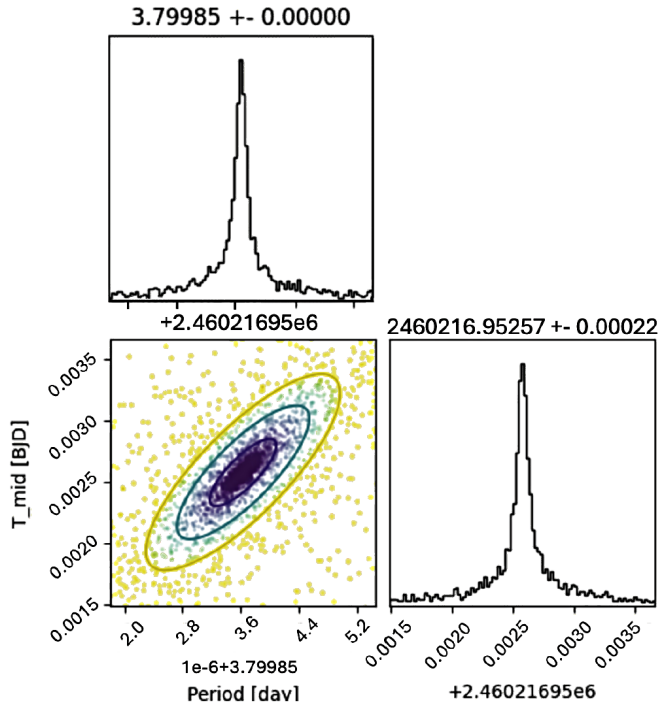

Figure 4. The posterior plot distribution for our newly calculated mid-transit time and orbital period using data from this study and previous studies.

the previously cited times to our new mid-transit time. The propagated mid-transit uncertainties were calculated using the following equation from Zellem *et al.* (2020):

$$\Delta T_{mid} = (n_{orbit}^2 \cdot \Delta P^2 + 2n_{orbit} \cdot \Delta P \, \Delta T_0 + \Delta T_0^2)^{1/2} \quad (1)$$

The second term in Equation 2 was dropped since none of the previous publications reported their covariance term (Zellem *et al.* 2020). Negating the covariance term led to the forward-propagated mid-transit uncertainties to be slightly underestimated. The results of the propagations of the mid-transit uncertainties from Bakos *et al.* (2015) and Ivshina and Winn (2022) are shown in Table 3.

When comparing our mid-transit uncertainty to the propagated, previously published mid-transit uncertainties, we found that we decreased the mid-transit uncertainty by 98.43% since HAT-P-54 b's discovery in 2015 (Bakos *et al.* 2015). More notable is the decrease in the mid-transit uncertainty from Ivshina and Winn (2022) by 70.27%. We also compared our updated period uncertainty to those previously reported. We found that we decreased the uncertainty in the period by 97.36% since Bakos *et al.* (2015) and by 78.24% when compared to Ivshina and Winn (2022).

In the interest of viewing how our study has an impact on the future of HAT-P-54 b observations, we forward-propagated the mid-transit uncertainties from Bakos *et al.* (2015) and Ivshina and Winn (2022), as well as our reported mid-transit uncertainty to a potential future observation. While there are many possible studies HAT-P-54 b could be a part of in the future, we chose to look at Habitable Exoplanet Observatory (HabEx), which is currently projected to be launched by the year 2039 (NASA JPL 2023). Since we cannot be certain when it will be launched and operational, we chose 31 December 2039 at midnight (2466153.50000 JD) as a baseline time to propagate to. Assuming there are no transit timing variations present in the system (of which we find no evidence in this study), we used Equation 1 to calculate the propagated mid-transit times (Zellem *et al.* 2020). After calculating and comparing each study's propagated uncertainty, we arrived at the values found in Table 4. By the launch date, our mid-transit uncertainty would be 98.28% less than that of Bakos *et al.* (2015) and 81.21% less than that of Ivshina and Winn (2022). This highlights the importance of frequent, ground-based follow-up analyses of exoplanets such as HAT-P-54 b.

Table 3. Updated ephemerides of HAT-P-54 b.

| Mid-transit (BJD_TBD) | Mid-transit Uncertainty (days) | Propagated Mid-transit Uncertainty (days) | Period (days) | Period Uncertainty (days) | Reference |
|---|---|---|---|---|---|
| 2460216.95257 | 0.00022 | N/A | 3.79985363 | 0.00000037 | This Work |
| 2456299.30370 | 0.00024 | 0.014 | 3.799847 | 0.000014 | Bakos *et al.* (2015) |
| 2458864.20475 | 0.00042 | 0.00074 | 3.7998529 | 0.0000017 | Ivshina and Winn (2022) |

Table 4. Propagated uncertainties to 2039-12-31 and the percent our study improved them by.

| Original Mid-Transit Uncertainty (d) | Propagated Mid-Transit Uncertainty (d) | Improved (%) | Reference |
|---|---|---|---|
| 0.00024 | 0.036 | 98.28 | Bakos *et al.* (2015) |
| 0.00042 | 0.0033 | 81.21 | Ivshina and Winn (2022) |
| 0.00022 | 0.00062 | N/A | Our data |

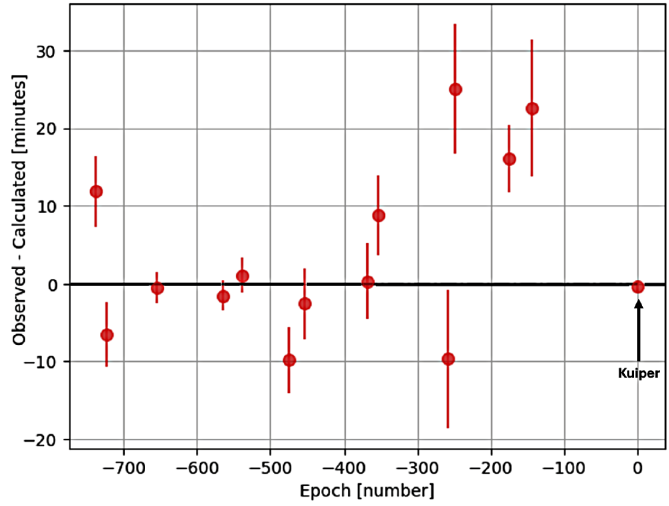

Figure 5. O–C plot for HAT-P-54 b using $t_0$ = 2460216.95298 and p = 3.79985392 as priors. The Kuiper Observation is marked with an arrow to distinguish it from the MicroObservatory Observations.

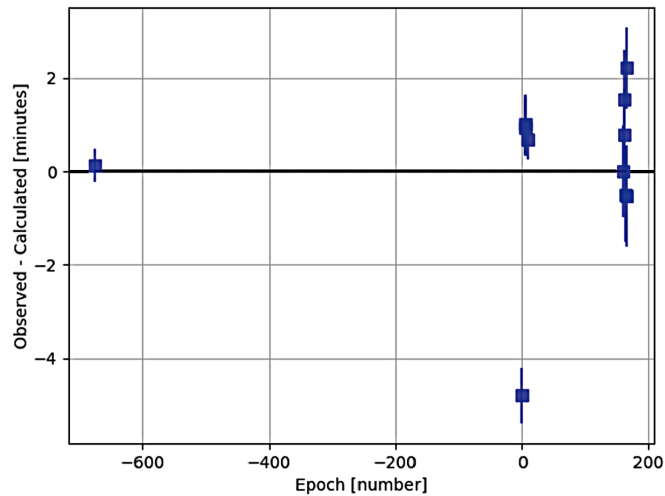

Figure 6. O–C plot using only previously published data and $t_0$ = 2459494.97987 and p = 3.79985392 as priors.

In addition to Figure 3, we created two additional O–C plots to compare the updated mid-transit time and orbital period we calculated (by leveraging a combination of this study's data and previously published data) to ephemeris values obtained from (1) this study's data exclusively, and (2) previously published data alone. Figure 5 presents an O–C plot that includes only the 14 data points from this study (shown in Table 1). Similar to the O–C plot presented in Figure 3, the mid-transit time from our most recent observation and the most recently published period, from Ivshina and Winn (2022), were used as priors. Alternatively, Figure 6 presents an O–C plot using only the previously published data reported in Table 2. For this plot, the mid-transit time from Ivshina and Winn's (2022) most recent observation (2459494.97987 ± 0.0007512) and their reported period (3.7998529 ± 0.0000017) were used as priors. The posterior plot distributions corresponding to these O–C plots are presented in Figures 7 and 8. Using the ephemeris fitter with only the 14 data points obtained from this study, we calculated an updated mid-transit time of 2460216.95338 ± 0.00044 BJD_TDB and an updated orbital period of 3.79985662 ± 0.0000014 days. Using only the data found from previous studies (Bakos *et al.* 2015; Saha *et al.* 2021; Ivshina and Winn 2022), we found the mid-transit time to be 2459494.98024 ± 0.00019 BJD_TDB and the orbital period to be 3.79985331 ± 0.00000038 days. Again, in order to compare the mid-transit uncertainties accurately, it is necessary to propagate these newly calculated mid-transit times to our original result that includes both this study's data and professional data (2460216.95257 BJD_TDB). In addition to this comparison, it is important to propagate all three mid-transit times to a future date to show the robustness of the different measurements over time. Therefore, we also propagated each of the derived mid-transit time values to the date mentioned previously (31 December 2039; 2466153.50000 JD). The results are presented in Table 5.

Although our mid-transit time uncertainty derived from a combination of this study's data and previously published data is slightly larger (9%) than the uncertainty derived from previously published data exclusively, it is apparent that our updated mid-transit time uncertainty lends itself to a more robust solution over time. When each of the mid-transit times is propagated to 31 December 2039, we find that our combined data mid-transit time uncertainty is 10% smaller than the uncertainty calculated using previously published data alone. Additionally, although the mid-transit time and orbital period values reported from the use of the combined data from this study and previous studies are more robust, it is important to note that the values obtained using the data from this study alone were still capable of producing a mid-transit time and period with reasonable uncertainties.

Table 5. Propagations of each calculated mid-transit time to our reported mid-transit time and 2039-12-31.

| Data Group | Original Mid-Transit Uncertainty (days) | Propagated Mid-Transit Uncertainty to 2460216.95257 (days) | Propagated Mid-Transit Uncertainty to 2039-12-31 (days) |
| --- | --- | --- | --- |
| Combined data | 0.00022 | N/A | 0.00062 |
| Data from this study only | 0.00044 | N/A | 0.0022 |
| Previously published data only | 0.00019 | 0.00020 | 0.00069 |

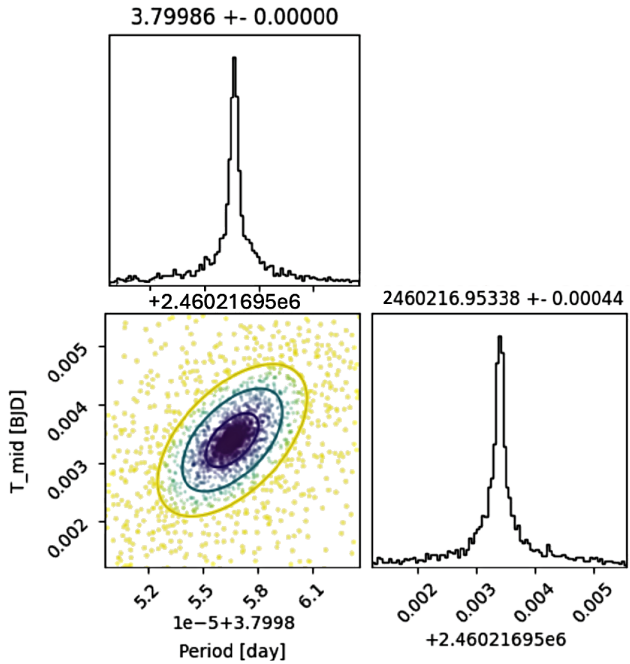

Figure 7. The posterior plot distribution for the calculated mid-transit time and orbital period using only data from this study.

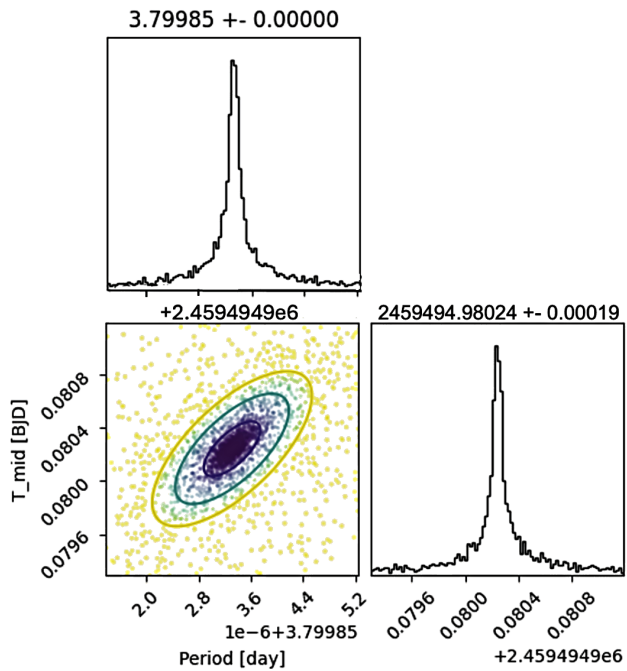

Figure 8. The posterior plot distribution for the calculated mid-transit time and orbital period using only previously published data.

### 6. Conclusions

In this study we demonstrate that by empowering a group of online, undergraduate students with the appropriate training and resources, citizen science is a viable option for the maintenance of exoplanet ephemerides. Leveraging data obtained from ground-based telescopes combined with data from previous studies, we were able to further reduce the orbital period and mid-transit time uncertainties for HAT-P-54 b. Most strikingly, our results showed a 70.27% improvement in the mid-transit uncertainty when compared to the most recent ephemeris update conducted by Ivshina and Winn (2022). Additionally, we found that using a combination of data from this study and previously published studies to update the mid-transit time provided the most robust solution over time. This finding in particular reaffirms the potential that citizen scientists have in contributing to the study of exoplanets, while also emphasizing the importance of ongoing efforts to improve the accuracy of exoplanet ephemerides in order to aid space-based telescopes in the efficient scheduling of observations. It demonstrates the power of leveraging both amateur and professional data.

This study and several others in the recent past (e.g. Zellem *et al.* 2020; Mizrachi *et al.* 2021; Hewitt *et al.* 2023a), prove that the ever-increasing number of exoplanet discoveries fosters an opportunity for collaboration between professional and amateur astronomers via citizen science projects. It confirms that small observatories and contributions from hobbyist equipment can have a direct positive impact on professional budgets for both time and money associated with projects like JWST and other large observatories through consistent and accurate updates to ephemerides over time.

The work completed in this study was done as a part of one of the first online Course-based Undergraduate Research Experiences (CUREs) for astronomy majors. CUREs make authentic research experiences accessible to a more diverse learning population and this study is a prime example of the importance of the work done by undergraduate students. A CURE's focus on scientific practices, discovery, collaboration, and authentic research (Auchincloss *et al.* 2014) make studies like this possible during a 15-week online course. The development and assessment of this CURE is presented in Hewitt *et al.* (2023b).

## 7. Acknowledgements

These observations were conducted with MicroObservatory, a robotic telescope network maintained and operated as an educational service by the Center for Astrophysics | Harvard & Smithsonian. MicroObservatory is supported by NASA's Universe of Learning under NASA award number NNX16AC65A to the Space Telescope Science Institute.

This publication makes use of data products from Exoplanet Watch, a citizen science project managed by NASA's Jet Propulsion Laboratory on behalf of NASA's Universe of Learning. This work is supported by NASA under award number NNX16AC65A to the Space Telescope Science Institute, in partnership with Caltech/IPAC, Center for Astrophysics | Harvard & Smithsonian, and NASA Jet Propulsion Laboratory.

This research has made use of the NASA Exoplanet Archive, which is operated by the California Institute of Technology, under contract with the National Aeronautics and Space Administration under the Exoplanet Exploration Program.

This work is supported by the National Science Foundation (NSF) under grant #IUSE 2121225.

Part of the research was carried out at the Jet Propulsion Laboratory, California Institute of Technology, under a contract with the National Aeronautics and Space Administration (80NM0018D0004).

---

[3] Univ. Arizona (2023). http://james.as.arizona.edu/~psmith/61inch/CCD/basicinfo.html

**Appendix A: Significant Detections of HAT-P-54 b**

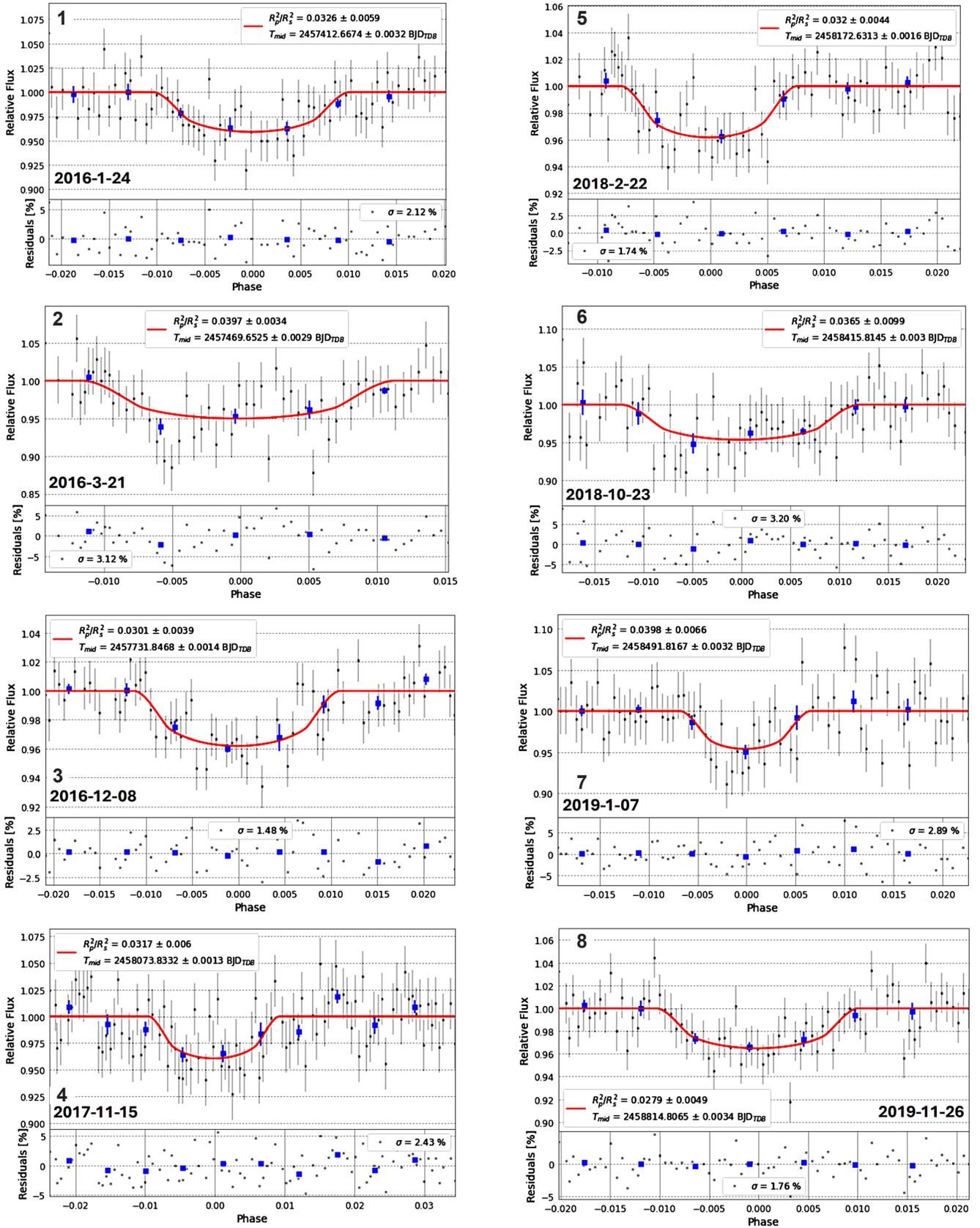

Figure A1. Light curves from this study. Light curves 1 through 13 were obtained using MicroObservatory and light curve 14 was obtained using the 61-inch Kuiper Telescope.

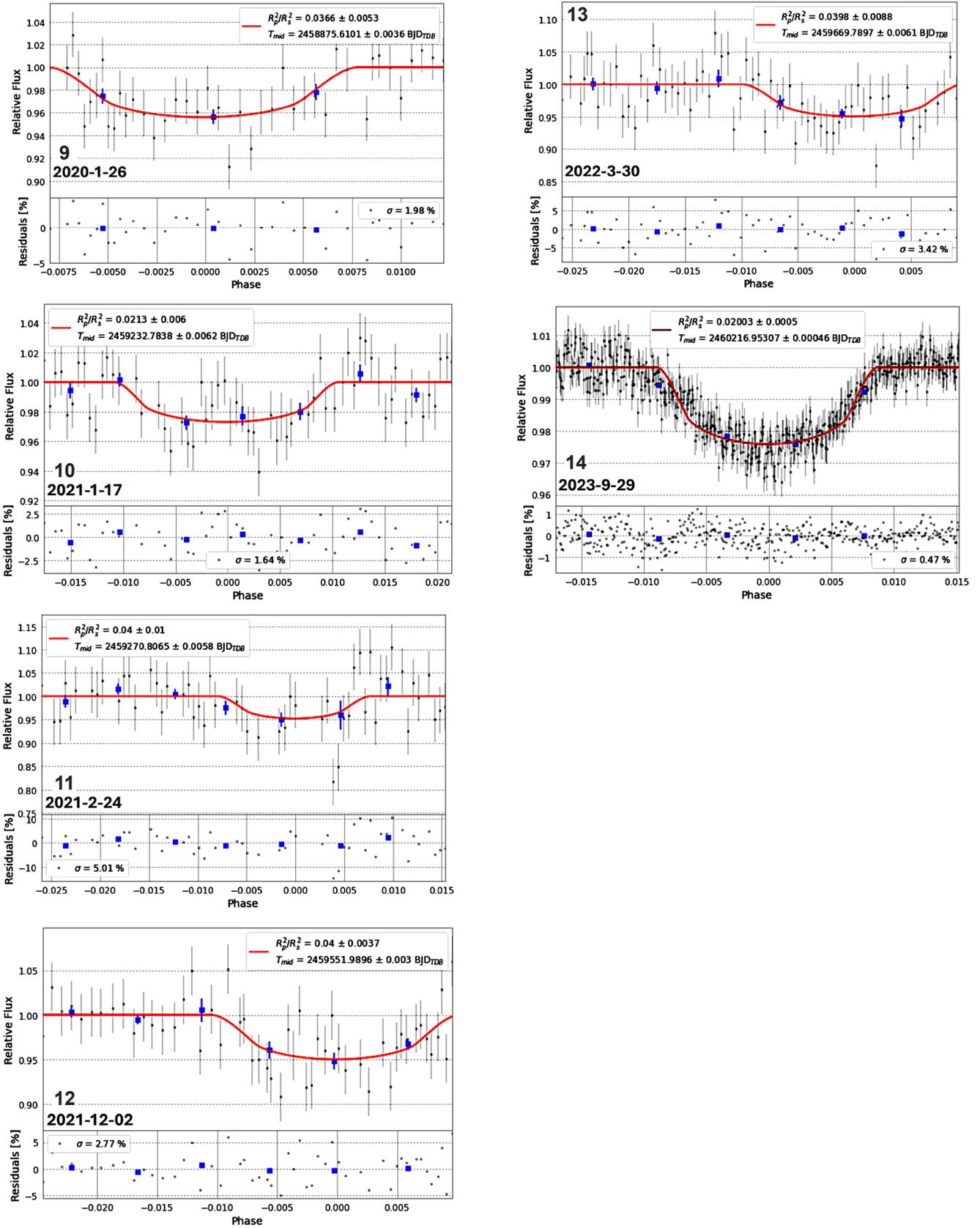

Figure A1. Light curves from this study, cont.